\documentclass{evn2004}
\usepackage{graphicx}
\begin{document}
   \title{Distance of W3(OH) by VLBI annual parallax measurement}

   \author{K. Hachisuka\inst{1}
          \and
          A. Brunthaler\inst{2,1}
          \and
          Y. Hagiwara\inst{3}
          \and
	  K. M. Menten\inst{1}
	   \and
          H. Imai\inst{4}
          \and
          M. Miyoshi\inst{5}   
          \and
          T. Sasao\inst{6}
          }

   \institute{Max-Planck-Institut f\"ur Radioastronomie, Auf dem H\"ugel 69,
              53121 Bonn, Germany
	      \and
	      Joint Institute for VLBI in Europe, Postbus 2, 
              7990 AA Dwingeloo, The Netherlands
	      \and
	      ASTRON, Westerbork Observatory, P.O. Box 2, 
	      7990 AA Dwingeloo, The Netherlands
	      \and
	      Department of Physics, Kagoshima University, Kagoshima 
              890-0065, Japan
	      \and
	      National Astronomical Observatory, Mitaka, Tokyo 181-8588, Japan
              \and 
	      Department of Space Survey and Information Technology, 
              Ajou University, Suwon, 442-749, Republic of Korea
             }

   \abstract{
The most powerful tool for measuring distances within our Galaxy is the 
annual parallax. We carried out phase-referencing VLBI observations of 
H$_{2}$O masers in the star forming region W3(OH) with respect to the 
extragalactic continuum source ICRF 0244+624 to measure their absolute 
proper motions. The measured annual parallax is 0.484 $\pm$ 0.004 
milli-arcseconds which corresponds to a distance of 
2.07$\mbox{}^{+0.01}_{-0.02}$ kpc from the sun. This distance is 
consistent with photometric and kinematic distances from previous 
observations.
}

   \maketitle
%
\section{Introduction}
The annual parallax is the most powerful tool for the determination 
of distances to objects in our Galaxy. The Hipparcos satellite successfully
measured the distances to many stars in the Solar neighborhood.
Those results contributed to various field of 
modern astronomy (e.g. Perryman et al. 1995). However, annual parallax 
measurements for far-away sources with kpc distances using single 
telescopes are impossible, 
since the value of the annual parallax is very small, 
e.g., 1 milli arcsecond (mas) at a distance of 1 kilopersecs (kpc). 
In fact, the Hipparcos satellite only determined the 
distances of stars within 200 pc from the Sun.

On the other hand, Very Long Baseline Interferometry (VLBI) provides the 
highest resolution in astronomy. In phase-referencing VLBI, the position of 
a target source is measured relative to 
a nearby positional reference 
source  (see e.g. Beasley \& Conway 1995 ; Ros 2003).  
The feasibility of annual parallax measurements with VLBI has been 
demonstrated by Brisken et al. (2002) who measured annual parallaxes of 
pulsars in the Galaxy and by van Langevelde et al. (2000) and  Vlemmings 
et al. (2002) who measured distances of Galactic OH masers associated with 
late type stars. Their results indicate that 
VLBI astrometry can measure distances of up to a few kpc. Hence, 
a large part of the Milky Way can be accessed by VLBI. This enables us to 
understand the Galactic structure and dynamics by VLBI astrometry since many 
maser sources exist in the whole Galaxy (e.g. Wouterloot et al. 1993) 
and are high brightness VLBI targets.

As our first step into this field of  study, we observed H$_{2}$O 
masers in the
Galactic star forming region W3(OH) and, alternately,  the close-by
extragalactic continuum 
source J0244+624 with phase-referencing VLBI. Here we report the result of an 
annual parallax measurement of H$_{2}$O maser features in W3(OH).

\section{Observation and data reduction}

\subsection{Observed sources}
We selected W3(OH) as our target source since it is one of the strongest 
Galactic H$_{2}$O maser sources. Located in the Perseus arm, W3(OH) 
was thought to have a distance of about 
2.3 kpc from the Sun (Georgelin \& Georgelin 1976; Humphreys 1978). 
\textit{Relative} proper motions 
of H$_{2}$O masers in W3(OH) 
have been measured before with VLBI (Alcolea et al. 1992). These masers 
move in a bipolar outflow, originating in the 
so-called Turner-Welch (TW) object, which is thought to be a high or 
intermediate-mass protostar (e.g., Reid et al. 1995; Wyrowski et al. 1999). 

We used ICRF 0244+624 as a phase-reference source. Its angular separation 
from W3(OH) is 2.17 degrees. Since this source is extragalactic with a 
redshift of 0.0438 (Margon \& Kwitter 1978), its own proper motion is 
negligible. It is also very strong and compact at lower frequencies 
(Fey \& Charlot. 2000). 
Thus, it is very useful as a phase-reference source.

\subsection{VLBA multi-epoch observations}
Seven epochs of observations were made over the course of 
16 months (Table 1). Time separations were from two to four months.
Each observation was carried out during 4 hours including 
calibrator observations.  
NRAO150 was observed for 5 minutes every 44 minutes for delay and 
bandpass calibration. W3(OH) and ICRF 0244+624 were observed with fast 
anntena-switching phase-referencing mode and at high elevation. 
The typical elevation varied from 46 to 62 degrees. The switching cycle 
was 40 seconds at all epochs, typical on-source time was 7 seconds for 
W3(OH) and ICRF 0244+624, respectively.

All observations were made with the ten station NRAO VLBA\footnote{The 
National Radio Astronomy Observatory is a facility of the National 
Science Foundation operated under cooperative agreement by Associated 
Universities, Inc.}. Two antennas (KP and LA) did not observe in the first 
epoch because of heavy snow, one anntena (PT) was flagged in the third and 
seventh epoch since most of the data was lost because of system troubles.
Also, SC was flagged in all epochs, since it produced no useful data.  
All Data were recorded using the VLBA recorder with two base-band 
channels (BBCs) and a bandwidth of 16 MHz. Data correlation was made with 1024 
spectral channels in each BBC with an integration time of 2 seconds. 
The resulting velocity spacing of each  spectral channel was 0.224 km s$^{-1}$ 
at 22.24 GHz. 

\begin{table}[htbp]
\caption[]{Summary of the VLBA observations.} 
\centering
\begin{footnotesize}
\begin{tabular}{ccc}
\hline 
\multicolumn{2}{c}{Epoch} & Stations \\ \hline
2001/01/28&01:11:00 -- 05:13:00 (UT)& 7 \\
2001/05/12&18:20:00 -- 22:20:00 (UT)& 9 \\
2001/07/12&14:20:00 -- 18:20:00 (UT)& 8 \\
2001/08/25&11:27:00 -- 15:27:00 (UT)& 9 \\
2001/10/23&07:35:00 -- 11:35:00 (UT)& 9 \\
2002/01/12&02:16:00 -- 06:16:00 (UT)& 9 \\
2002/05/06&18:44:00 -- 22:44:00 (UT)& 7 \\
\hline
\end{tabular}
\end{footnotesize}
\end{table} 

\subsection{Data reduction}
The data were calibrated and imaged with standard techniques using
the AIPS software package. A priori amplitude calibration was applied using
system temperature measurements and standard gain curves. We calibrated clock 
offsets between antennas using the calibrator NRAO150. A fringe fit was
performed on ICRF 0244+624 and the solutions were applied to W3(OH).

The largest errors in phase-referencing observations are introduced by a 
zenith delay error in the atmospheric model of the correlator. These errors 
will degrade the image quality and the astrometric accuracy. However, we
 modeled the calibrated phase data of the strongest maser feature as the 
result of an position offset and a zenith delay error at each station. 
These zenith delay errors can then be corrected by the AIPS task CLCOR. 
This correction improves the quality of the phase-referenced images and 
the astrometric accuracy (e.g. Brunthaler et al. 2003).
The data of MK at the 7th  epoch was flagged since we 
could not estimate the zenith delay error of this station. 

After those 
corrections, we carried out fringe fitting for NRAO150 and ICRF 0244+624 
again. Finally, we applied these revised solutions to the UV data of W3(OH) 
and created images of the maser features at each epoch. 

To determine the position of each maser spot, we performed a two-dimensional 
Gaussian fit with the AIPS task JMFIT. We traced maser features between 
different epochs by taking account of their LSR velocities and positions. 
It is difficult to exactly trace the same maser feature since H$_{2}$O 
masers are highly time variable and their absolute proper motions relative 
to the extragalactic reference source were non-linear because of the effect 
of the parallax. 

\section{Results}

\subsection{Reference source: ICRF 0244+624}
The extragalactic continuum source ICRF 0244+624 was detected in all epochs.
The peak flux densities were always over 0.9 Jy/beam, which is strong enough 
for a excellent phase calibration. The structure of this source was compact 
and unresolved like in previous VLBI observations at lower frequencies. 
Positions of the peak between all epochs were fluctuating within  
$\pm 10 \mu$as from the map origin. The standard deviations of the peak 
position were 6.5 $\mu$as in right ascension and 7.6 $\mu$as in declination, 
which we take as  the uncertainty of the reference position.  

\subsection{Galactic H$_{2}$O maser source: W3(OH)}
H$_{2}$O masers in W3(OH) were detected in all epochs.
The peak flux densities were from a few hundred mJy to one thousand Jy.
The masers were distributed over an area of 2.5'' $\times$ 0.5'' 
(Figure 1) which is consistent with previous VLBI observations 
(Alcolea et al. 1992). Figure 1 contains only the 15 maser features that were 
detected in at least 5 epochs. These features are listed in Table 2. 
The maser features were usually detected in 2 to 7 adjacent velocity channels. 
There were additionally many other short lived H$_{2}$O maser features. 

The typical formal position error of each H$_{2}$O maser spot was 
10 $\mu$as in right ascension  and 20 $\mu$as in declination. 
Figure 2 shows the absolute proper motion of the maser feature 2 in Table 2. 
This proper motion is the sum of the inner motion in W3(OH), the annual 
parallax, the Galactic rotation, Solar motion and probably a peculiar motion 
of W3(OH) which might differ from the Galactic rotation. All motions except 
the annual parallax are linear motions. Also, these motions are equal for 
all maser features except the inner motion. 

\begin{table}[htbp]
\caption[]{List of detected maser features. The columns give the name, 
number of channels with emission used in the model fit, 
number of epochs, position offset from the phase center at first 
detected epoch and LSR velocity.}
\centering
\begin{footnotesize}
\begin{tabular}{cccr@{}c@{}lr@{}c@{}lc}
\hline 
No. &  channels& epochs &
\multicolumn{6}{c}{Position offset}& V$\mbox{}_{LSR}$  \\
& & & \multicolumn{3}{c}{$\Delta \alpha$ (mas)}
    & \multicolumn{3}{c}{$\Delta \delta$ (mas)}
& (km s$\mbox{}^{-1}$)\\ \hline
 1 &4 &7 &   --4&.&99 &  --5&.&6 & --48.6\\ 
 2 &4 &7 &   --9&.&68 &   33&.&4 & --49.0\\ 
 3 &3 &6 &   --7&.&72 &   33&.&3 & --48.6\\ 
 4 &2 &5 &   --5&.&62 &   32&.&9 & --50.1\\ 
 5 &3 &5 &  --13&.&70 &   49&.&2 & --49.4\\ 
 6 &2 &6 &  --21&.&68 &   63&.&2 & --49.0\\ 
 7 &3 &5 &  --19&.&54 &   62&.&2 & --48.8\\ 
 8 &4 &6 &  --38&.&16 &   97&.&9 & --49.4\\ 
 9 &7 &7 & --144&.&50 &--144&.&3 & --51.1\\ 
10 &4 &5 & --653&.&64 &   92&.&1 & --52.6\\ 
11 &2 &6 & --646&.&07 &   90&.&2 & --51.5\\ 
12 &4 &5 & --653&.&27 &   92&.&3 & --51.9\\ 
13 &3 &6 & --662&.&86 &   95&.&0 & --54.7\\ 
14 &5 &7 &--2045&.&84 &--133&.&4 & --45.2\\ 
15 &5 &6 &--2231&.&28 &   84&.&1 & --62.8\\ 
\hline
\end{tabular}
\end{footnotesize}
\end{table} 

\begin{figure}[htbp]
   \centering
   \includegraphics[width=8cm]{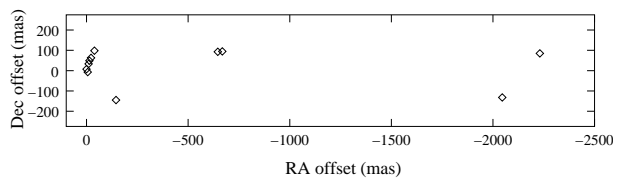}
   \caption{Spatial distribution of H$_{2}$O maser features in W3(OH). 
            The open square shows that the H$_{2}$O maser feature was 
	    detected over 5 epochs. 
	    \label{}
           }
\end{figure}

\begin{figure}[htbp]
   \centering
   \includegraphics[width=8cm]{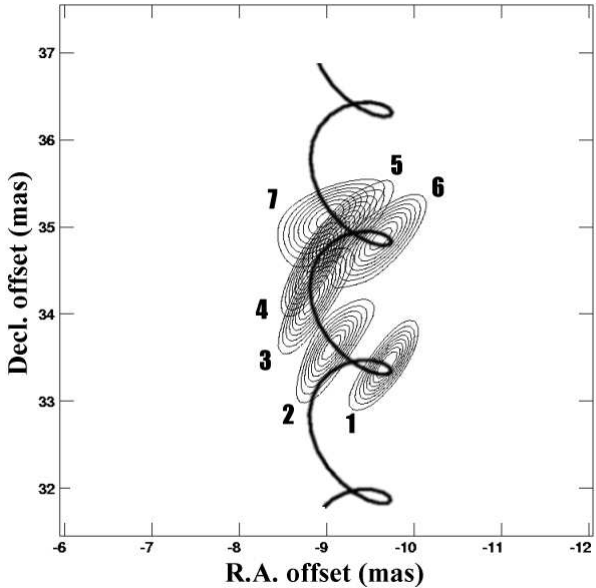}
   \caption{Absolute proper motion of the H$_{2}$O maser feature at 
     --49.0 km$\mbox{}^{-1}$ (No. 2 in Table 2) in the image 
     plane. The numbers indicate each epoch. Contours are 20, 30, 
     40, 50, 60, 70, 80, 90\% of the peak of brightness. 
     The peak flux density of 1st to 7th epoch 
     was 430, 215, 204, 79, 31, 39, and 44 Jy/beam, respectively. 
     Solid line shows the result of astrometric model fitting. 
            \label{}
           }
\end{figure}

\begin{figure}[htbp]
   \centering
   \includegraphics[width=8cm]{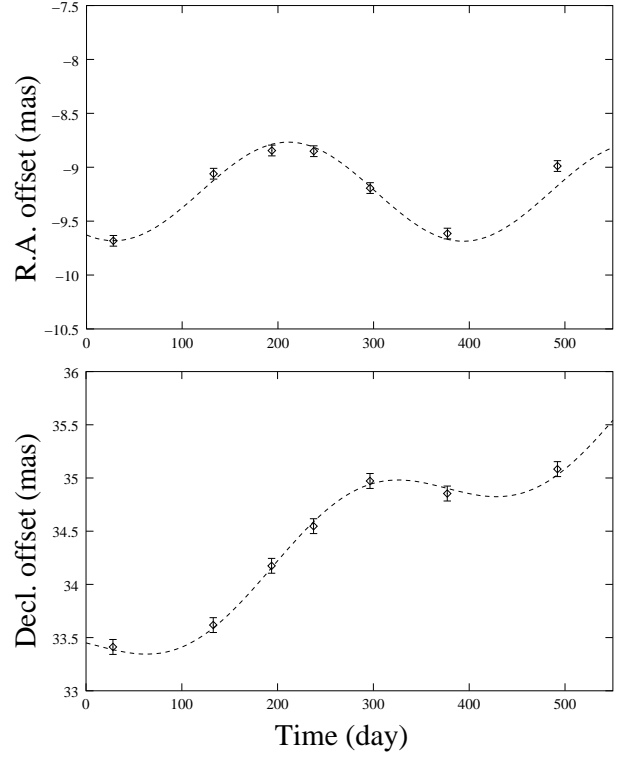}
   \caption{Absolute proper motion of H$_{2}$O maser at LSR velocity of  
     --49.0 km$\mbox{}^{-1}$ (No. 2 in Table 2.) at  
     right ascension and declination, respectively.
     The dashed line shows the best fit curve by the astrometric model 
     fitting.
            \label{}
           }
\end{figure}

\subsection{Annual parallax of maser features}

We carried out a model fit of the path of the maser features 
in terms of its proper motion ($\mu_{\alpha},\mu_{\delta}$) and 
annual parallax ($\Pi$) by
\begin{eqnarray*}
\Delta \alpha \cos \delta & = & \Pi f_{\alpha}(\alpha,\delta,t) 
+ \mu_{\alpha} t + \alpha_{0} \\
\Delta \delta & = & \Pi f_{\delta}(\alpha,\delta,t) 
+ \mu_{\delta} t + \delta_{0} 
\end{eqnarray*}
where t is time.
The functions $f_{\alpha}$ and $f_{\delta}$ are the parallax 
displacements in right ascension and declination, respectively.
The origin of position for each maser feature, $\alpha_{0}$ and 
$\delta_{0}$, is not discussed here. 

Maser features 14 and 15 are located 2 arcseconds from the phase center 
and are affected by time-smearing effects. Hence, they were excluded from our 
analysis. We used a total of 45 different data sets (13 maser features 
in several 
velocity channels). We fit one annual parallax and 45 individual linear 
motions to the combined data set. We ignored any acceleration of maser 
features in the proper motions and radial velocities. The estimated annual 
parallax is 0.484 $\pm$ 0.004 mas. Figure 3 shows the result of the 
astrometric model fitting for one channel of maser feature 2. 

We also fitted a proper motion and an annual parallax to each of the 45 data 
sets individually. The errors of these individual parallaxes range from $\sim$ 
0.01 mas for strong components to $\sim$ 0.1 mas for weak components. 
The parallaxes of the different maser features were in general consistent 
within the errors with the parallax from the combined fit. We can define 
a parameter R$_i$ for each individual fit by
\begin{eqnarray*}
R_i=\frac{|\pi_i-\bar\pi|}{\Delta\pi_i},
\end{eqnarray*}
where $\pi_i$ and $\Delta\pi_i$ are the parallax and its error from fit i 
and $\bar\pi$ is the parallax from the combined fit.

\begin{figure}[htbp]
   \centering
   \includegraphics[width=6cm,angle=-90]{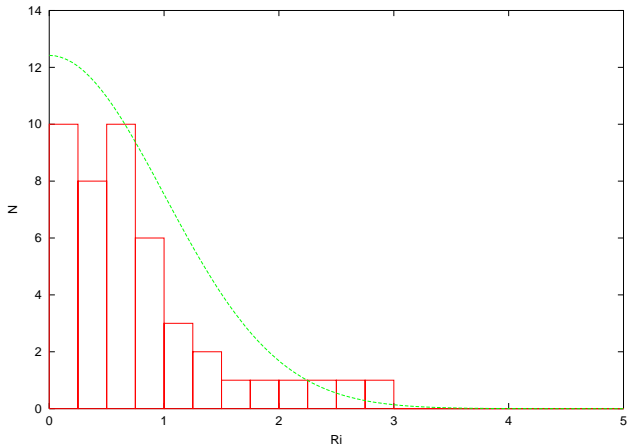}
   \caption{Histogram of parameter R$_i$ (see text) for the 
            individual fits and the expectation from a Gaussian
	    distribution of errors (dashed line).
            \label{}
           }
\end{figure}

A histogram of this parameter R$_i$ is shown in Figure 4 together with the 
expectation for a Gaussian distribution of parallaxes. It can be clearly seen 
that the data is in good agreement with a Gaussian distribution. Hence, 
the fitting results are dominated by statistical and not by systematic errors. 

An annual parallax for W3(OH) of 0.484 $\pm$ 0.004 kpc corresponds to a 
distance of 2.07$\mbox{}^{+0.01}_{-0.02}$ kpc. Georgelin \& Georgelin (1976) 
estimated a kinematic distance of 2.3 kpc for W3(OH) and  Humphreys (1978) 
estimated a photometric distance of 2.2 kpc to the OB association near W3(OH). 

\section{Conclusion}
We estimated an annual parallax of 0.484 $\pm$ 0.004 mas for W3(OH)
from the absolute proper motions of associated H$_{2}$O maser features.
This corresponds to a distance of 2.07$\mbox{}^{+0.01}_{-0.02}$ kpc 
and is consistent with past determinations. 

The absolute proper motion is the sum of the annual parallax and a linear 
component. The linear component includes the inner motion of each 
maser feature, which can be estimated independently from the motion 
relative to a reference maser feature. 
After subtracting the inner motions, all maser components will have the 
same proper motions. The linear part of these motions can be determined 
from maser features which are detected in more than five epochs.
Thus, we can decrease the  number of free parameters by fixing 
the linear component. Then
we can also use components which are detectable in less than 5 epochs.
This has the potential to increase the accuracy of the annual parallax even 
further. We will also get the total proper motion of W3(OH) from this 
observation and this will help to constrain models of the Galactic 
rotation. This will be the subject of an upcoming paper.

There are other H$_{2}$O of other maser sources with close-by positional 
reference sources in the Galaxy. Thus, it is possible to determine the 
distance of other maser sources with phase-referencing VLBI astrometric 
observations with similar accuracy to the case of the  H$_{2}$O masers 
in W3(OH). 
We are planning an extensive program to use parallax and 
proper motion measurements to constrain Galactic structure and rotation.

\begin{acknowledgements}
We are grateful to Dr. Mark Reid for very helpful discussions.
\end{acknowledgements}


\begin{thebibliography}{}
  \bibitem[1992]{Alcolea92} Alcolea, J., Menten, K. M., Moran, J. M., 
\& Reid, M. J. 1992, in Astrophysical Masers, ed. A. W. Clegg \& G. E. 
Nedoluha (Heidelberg:Springer), 225
  \bibitem[1995]{Beasley95} Beasley, A. J., \& Conway, J. E. 1995, 
Very Long Baseline Interferometry and the VLBA, ed. J. A. Zensus, P. J. 
Diamond, \& P. J. Napier (San Francisco: ASP), ASP Conf. Ser., 82, 328
  \bibitem[2002]{Brisken02} Brisken, W. F., Benson, J. M. \& Goss, W. M., 
2002, Ape, 571, 906
  \bibitem[2003]{Andreas03} Brunthaler, A., Reid, M. J. \& Falcke, H., 
2003, astro-ph/0309575
  \bibitem[2000]{Fey00} Fey, A. L. \& Charlot, P. 2000, ApJS, 128, 17
  \bibitem[1976]{Georgelin76}  Georgelin, Y. M., \& Georgelin, Y. P.
1976, A\&A, 49, 57 
  \bibitem[1978]{margon78} Margon, B. \& Kwitter, K. B. 1978, Apj, 224, L43
  \bibitem[1995]{Perryman95} Perryman, M. A. C., Lindegren, L., 
Kovalevsky, J., et al. 1995, A\&A, 304, 69
  \bibitem[1995]{Reid95} Reid, M. J., Argon, A. L., Masson, C. R., et al. 
1995, ApJ, 238, 443
\bibitem[2003]{Ros03} Ros, E. 2003, astro-ph/0308265
  \bibitem[2000]{vLangevelde00} van Langevelde, H. J., Vlemmings, W., 
Diamond, P. J., et al. 2000, A\&A, 357, 945
  \bibitem[2003]{Vlemmings03} Vlemmings, W., van Langevelde, H. J.,
Diamond, P. J., et al. 2003, A\&A, 407, 213
\bibitem[1993]{wbf93}
  Wouterloot, J.G.A., Brand, J., \& Fiegle, K. 1993, A\&AS, 98, 589
\bibitem[1999]{[Wyrowski99} 
Wyrowski, F., Schilke, P., Walmsley, C.~M., \& Menten, K.~M.\ 1999, ApJ, 
514, L43 

\end{thebibliography}
\end{document}